\documentclass[aps,prl,twocolumn,superscriptaddress,floatfix,longbibliography,a4paper]{revtex4-1}

\usepackage{amsmath, amssymb}
\usepackage{mathtools}
\usepackage{lipsum}
\usepackage{times}
\usepackage{graphicx}
\usepackage{wasysym}
\usepackage{bm}
\usepackage{hyperref}
\usepackage{xspace}
\usepackage{siunitx}
\usepackage{color}
\usepackage[labelsep=space,labelfont={color=myColor,bf}]{caption}
\usepackage[normalem]{ulem}
\usepackage[margin={1.7cm,2.5cm}]{geometry}

\definecolor{myColor}{rgb}{0.02,0.12,0.3}
\definecolor{myciteColor}{rgb}{0.39,0.7,0.89}
\definecolor{myCsColor}{rgb}{0.1,0.1,0.4}
\hypersetup{colorlinks=true, linkcolor=myColor, filecolor=myColor, urlcolor=myColor, citecolor=myColor, urlcolor=myColor}
\newcommand{\kB}{\ensuremath{k_{\text{B}}}\xspace}
\newcommand{\UD}{\ensuremath{U_{\text{D}}}}

\DeclareSIUnit{\nK}{\nano\kelvin}
\DeclareSIUnit{\aB}{\emph{a}_0}
\DeclareSIUnit{\G}{G}
\newcommand{\K}[1]{\ensuremath{{}^{#1}\text{K}}}
\renewcommand{\figurename}[1]{Fig.~}

\begin{document}

\title{Bidirectional dynamic scaling in an isolated Bose gas far from equilibrium}

\author{Jake A.~P. Glidden}
\affiliation{Cavendish Laboratory, University of Cambridge, J. J. Thomson Avenue, Cambridge CB3 0HE, United Kingdom}
\author{Christoph Eigen}
\affiliation{Cavendish Laboratory, University of Cambridge, J. J. Thomson Avenue, Cambridge CB3 0HE, United Kingdom}
\author{Lena H. Dogra}
\affiliation{Cavendish Laboratory, University of Cambridge, J. J. Thomson Avenue, Cambridge CB3 0HE, United Kingdom}
\author{Timon A. Hilker}
\affiliation{Cavendish Laboratory, University of Cambridge, J. J. Thomson Avenue, Cambridge CB3 0HE, United Kingdom}
\author{Robert P. Smith}
\affiliation{Clarendon Laboratory, University of Oxford, Parks Road, Oxford OX1 3PU, United Kingdom}
\author{Zoran Hadzibabic}
\affiliation{Cavendish Laboratory, University of Cambridge, J. J. Thomson Avenue, Cambridge CB3 0HE, United Kingdom}


\maketitle

\newpage


\textbf{ 
Understanding and classifying nonequilibrium many-body phenomena, analogous to the classification of equilibrium states of matter into universality classes~\cite{Hohenberg:1977,Chaikin:1995}, is an outstanding problem in physics. 
Any many-body system, from stellar matter to financial markets, can be out of equilibrium in a myriad of ways; since many
are also difficult to experiment on, it is a major goal to establish universal principles that apply to different phenomena and physical systems. 
At the heart of the classification of equilibrium states is the universality seen in the self-similar spatial scaling of systems close to phase transitions. Recent theoretical work~\cite{Micha:2004, Berges:2008, Nowak:2012, Nowak:2014, Berges:2015b, PineiroOrioli:2015, Chantesana:2019, Mikheev:2019, Schmied:2019b, Bhattacharyya:2019, Berges:2019, Fujimoto:2020}, and first experimental evidence~\cite{Pruefer:2018, Erne:2018}, suggest that isolated many-body systems far from equilibrium generically exhibit dynamic (spatiotemporal) self-similar scaling, akin to turbulent cascades~\cite{Zakharov:1992} and the Family--Vicsek scaling in classical surface growth~\cite{Family:1985,Kardar:1986}.
Here we observe bidirectional dynamic scaling in an isolated quench-cooled atomic Bose gas; as the gas thermalises and undergoes Bose--Einstein condensation, it shows self-similar net flows of particles towards the infrared (smaller momenta) and energy towards the ultraviolet (smaller lengthscales).
For both infrared (IR) and ultraviolet (UV) dynamics we find that the scaling exponents are independent of the strength of the interparticle interactions that drive the thermalisation.
}

A key question in the quest to understand nonequilibrium dynamics is how an isolated quantum many-body system that is initially far from equilibrium thermalises~\cite{Polkovnikov:2011, Gogolin:2016} \footnote{Experiments on ultracold atomic gases~\cite{Langen:2015} have investigated various aspects of this problem, through studies of, {\it e.g.}, integrability~\cite{Kinoshita:2006}, prethermalization~\cite{Gring:2012, Eigen:2018}, generalised Gibbs ensembles~\cite{Langen:2014},  and the Eigenstate Thermalisation Hypothesis~\cite{Kaufman:2016}}.
Possible universal features of this process have recently been conceptualised in the framework of so-called non-thermal fixed points (NTFPs)~\cite{Berges:2008}, with spatiotemporal scaling predicted to occur in ultracold atomic gases~\cite{PineiroOrioli:2015, Berges:2015b, Chantesana:2019, Mikheev:2019}, quantum magnets~\cite{Bhattacharyya:2019}, and the quark-gluon plasma~\cite{PineiroOrioli:2015, Berges:2015b, Berges:2019}.
In the context of ultracold atoms, these theories give a new perspective to the foundational~\cite{Snoke:1989, Stoof:1991, Svistunov:1991, Kagan:1992, Semikoz:1995, Kagan:1996a, Damle:1996, Gardiner:1997a, Berloff:2002, Miesner:1998b, Kohl:2002} and still open~\cite{Davis:2017,Beugnon:2017} problem of the formation of a Bose--Einstein condensate (BEC).
First experimental evidence for the emerging NTFP paradigm was seen in one-dimensional (1D) harmonically trapped atomic gases~\cite{Pruefer:2018, Erne:2018}. So far, self-similar scaling was observed only in the IR dynamics. Our experiments reveal both IR and UV spatiotemporal scaling in the textbook setting of a homogeneous 3D Bose gas~\cite{Gaunt:2013}, with tuneable interactions
and near-perfect isolation from the environment.


\begin{figure*}
  \centering
  \includegraphics[width=\textwidth]{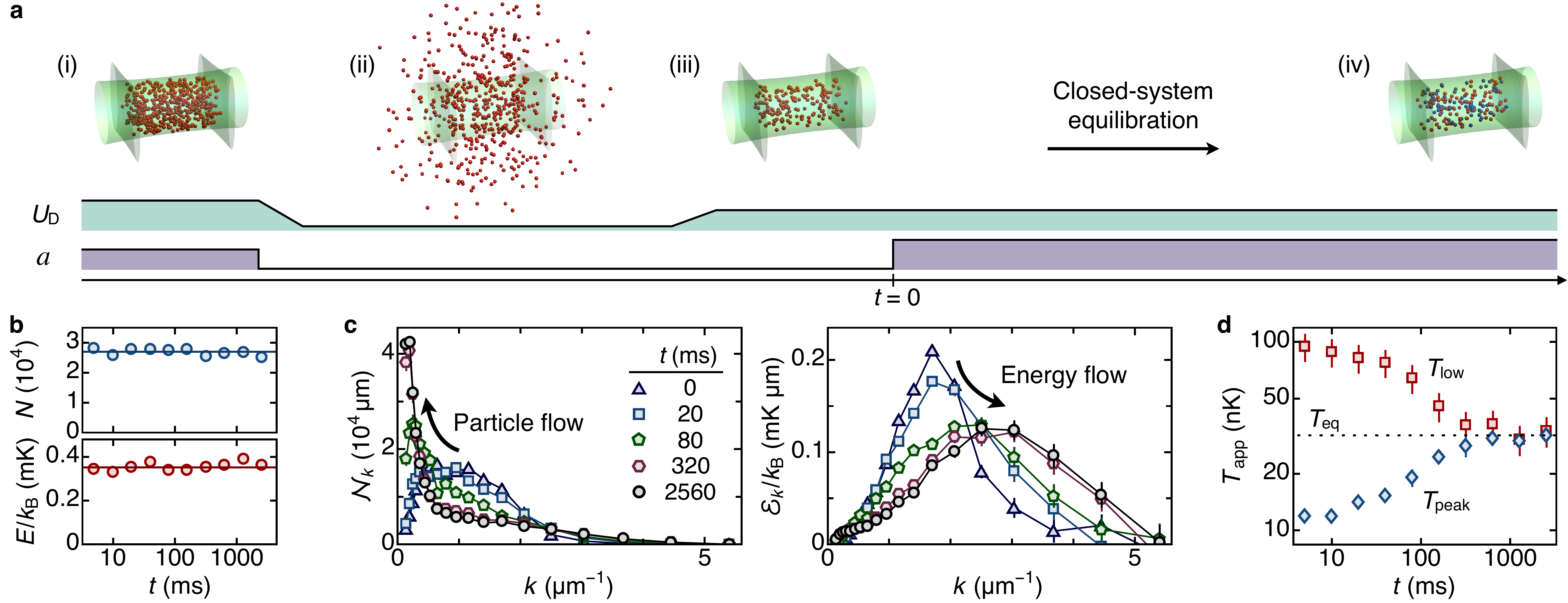}
 \caption{$\vert$ {\bf Bidirectional thermalisation in an isolated gas.} 
 {\bf a,} Experimental protocol; $\UD$ is the depth of the optical box trap (green) and $a$ the scattering length characterising interactions. (i)~We prepare an equilibrium cloud of $N\approx\num{1.2e5}$ atoms at $T\approx\SI{130}{\nK}$, just above the condensation temperature. (ii)~We turn off the interactions ($a\rightarrow 0$) and then lower $\UD$, so high energy atoms escape without the remaining ones thermalising. (iii)~We close the system by raising $\UD$ and then initiate thermalisation, at $t=0$, by tuning $a$ to a nonzero value. (iv)~Equilibrium state, with both thermal (red) and condensed (blue) components. 
 {\bf b - d,} Thermalisation at $a=300~a_0$. {\bf b,}~Total atom number, $N$, and energy, $E$, remain constant. {\bf c,} Evolution of the spectral distributions of particles, $\mathcal{N}_k$, and energy, $\mathcal{E}_k$; the net particle flow is to the IR and the net energy flow to the UV. 
 {\bf d,} The low-$k$ behaviour of $\mathcal{E}_k$ and the momentum where $\mathcal{E}_k$ peaks show different apparent temperatures ($T_{\rm low}$ and $T_{\rm peak}$, respectively), which converge to the expected equilibrium temperature $T_{\rm eq}$ on similar timescales. }
\label{fig:1}
\end{figure*}

The idea of our experiments is depicted in Fig.~\ref{fig:1}a. We start with an equilibrium homogeneous \K{39} gas of \mbox{$N\approx\num{1.2e5}$} atoms in the lowest hyperfine ground state, confined in a cylindrical optical box of diameter $D\approx \SI{27}{\um}$, length $L\approx \SI{46}{\um}$, and depth \mbox{$\UD \approx \kB\times\SI{1}{\micro\kelvin}$}~\cite{Eigen:2016}. The tuneable interactions in our gas are characterised by the scattering length $a$. Initially $a=200~a_0$, where $a_0$ is the Bohr radius, and \mbox{$T\approx\SI{130}{\nK}$}, just above the condensation temperature $T_{\rm c}$. 
We then create a far-from-equilibrium cloud by removing \SI{77}{\percent} of the atoms and \SI{97.5}{\percent} of the total energy $E$, so the energy per particle drops by an order of magnitude, and in equilibrium the gas would be (partially) condensed.
Using tuneable interactions allows us to completely separate this quench from the subsequent equilibration. 
First, we switch off the interactions (tune $a \rightarrow 0$) and then lower $\UD$ to $\kB\times\SI{30}{\nano\kelvin}$ for $\SI{2}{\second}$, so high energy atoms escape without the remaining ones thermalising~\footnote{Note that $\UD$ caps the component of an atom's momentum perpendicular to any of the trap walls, and not its total momentum, so in the absence of collisions some atoms with energy exceeding $\UD$ remain in the trap.}; this results in a far-from-equilibrium momentum distribution $n_k$. Then, only after closing the system by raising $\UD$~\footnote{We raise $\UD$ to $\approx \kB\times\SI{400}{\nano\kelvin}$, sufficient to prevent evaporation while avoiding technical heating during thermalisation. Since the optical-box walls are not infinitely sharp~\cite{Gaunt:2013}, the effective $D$ and $L$ depend slightly on $\UD$ and $E$; during thermalisation $D = \SI{25(2)}{\um}$ and $L = \SI{42(2)}{\um}$.}, we turn on the interactions (within a few milliseconds) and thus start the clock for thermalisation. To probe the state of the gas after a variable relaxation time $t$, we turn off both the trap and the interactions, and infer $n_k(k, t)$ from absorption images taken after ballistic expansion of the cloud (see Methods).

As shown in Fig.~\ref{fig:1}b, during thermalisation (at $300\,a_0$) the total $N$ and $E$ remain constant. 
In Fig.~\ref{fig:1}c we plot both the spectral population density $\mathcal{N}_k=4\pi k^2 n_k$ (left) and the spectral energy density $\mathcal{E}_k= \mathcal{N}_k \hbar^2 k^2/(2m)$ (right); here the conserved $N$ and $E$, respectively, correspond to the areas under the curves. 
As indicated by the arrows, we observe bidirectional dynamics in momentum space~\cite{Svistunov:1991}: while the majority of atoms flows to the IR, where the condensate emerges, the energy, carried by a small fraction of atoms, flows to the UV. 
In Fig.~\ref{fig:1}d we plot two different apparent temperatures, $T_{\rm peak}$ and $T_{\rm low}$, both deduced from $\mathcal{E}_k$ by (incorrectly) assuming equilibrium. For an equilibrium gas, at $T \leq T_{\rm c}$, one can simply get $T$ from $k_{\rm peak}$, the momentum where $\mathcal{E}_k$ peaks, but alternatively one can consider only the low-$k$ states, where $\mathcal{E}_k \propto T k^2$ for $k\rightarrow 0$. Here, the apparent $T_{\rm peak} \propto k_{\rm peak}^2$  is initially far below the equilibrium temperature, $T_{\rm eq} \approx 32$~nK, corresponding to the conserved $E$. On the other hand, the low-$k$ $T_{\rm low}$ is initially far above $T_{\rm eq}$ (and close to the pre-quench temperature). The two apparent temperatures thus evolve in opposite directions, and we find that they converge  to $T_{\rm eq}$ on similar timescales.


\begin{figure*}
   \centering
      \includegraphics[width=\textwidth]{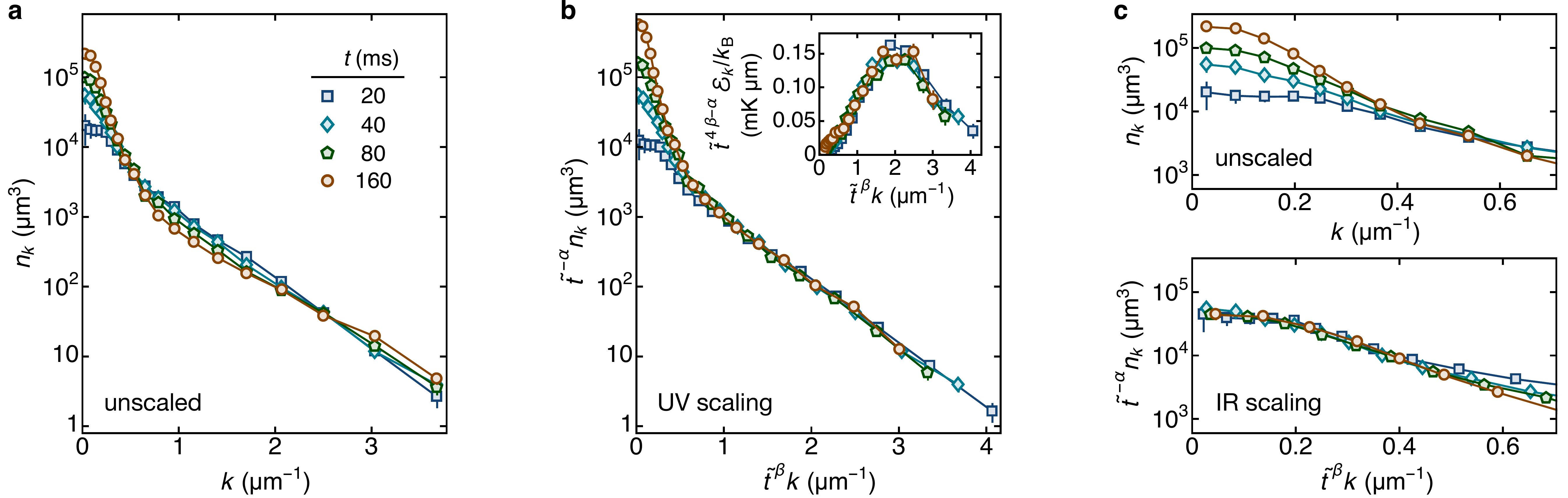}
   \caption{$\vert$ {\bf Self-similar scaling dynamics.} Here $a=300~a_0$ during closed-system thermalisation. For scaling $n_k$ according to Eq.~\eqref{eq:scaling}, we arbitrarily set $t_0 = 40\,$ms.
{\bf a,}~Unscaled $n_k$ curves; the legend applies to all panels. {\bf b,}~UV scaling, with $\alpha = -0.70(7)$ and $\beta = - 0.14(2)$, collapses the curves for $\tilde{t}^\beta k \gtrsim\SI{1}{\per\um}$.  The inset shows the scaled $\mathcal{E}_k$ curves, for comparison with Fig.~\ref{fig:1}c. Here $\alpha/\beta \approx 5$, consistent with energy-conserving transport. {\bf c,}~IR scaling. Top panel: a zoom-in on the unscaled $n_k$ curves at low $k$. Bottom panel: 
scaling with $\alpha = 1.15(8)$ and $\beta = 0.34(5)$ collapses the curves for $\tilde{t}^\beta k \lesssim\SI{0.5}{\per\um}$.
Now $\alpha/\beta \approx 3$, consistent with particle-conserving transport.}
  \label{fig:2}
\end{figure*}

The universal-scaling prediction is that at intermediate times, when the state of the system is distinct from both the initial and the final (equilibrium) one, the thermalisation dynamics can in some appropriate (IR and UV) momentum ranges be described by spatiotemporal scaling of the form
\begin{equation}
   n_k(\mathbf{k}, t)=\tilde{t}^{\alpha}\,n_k (\tilde{t}^{\beta}\mathbf{k},t_0) ,
   \label{eq:scaling}
\end{equation}
where  $t_0$ is a reference time, $\tilde{t}=t/t_0$, and the scaling exponents $\alpha$ and $\beta$ are positive (negative) for transport towards the IR (UV).  This implies that the particle and energy flows are akin to self-similar turbulent cascades~\cite{Dyachenko:1992, Navon:2016, Navon:2019} and that the $n_k$ distributions at different times can, separately in the IR and the UV, be collapsed onto universal curves.

 The bidirectional dynamics in our gas indeed show such spatiotemporal scaling. Specifically, for $a=300~a_0$ (as in Fig.~\ref{fig:1}), we observe dynamic scaling for $t \in [20~{\rm ms},160~{\rm ms}]$. In Fig.~\ref{fig:2}a we show the unscaled $n_k(t)$ curves.  For the scaled ones in Fig.~\ref{fig:2}b,c we arbitrarily set $t_0 = 40$~ms and have optimised their collapse by varying $\alpha$ and $\beta$ (see Methods).

In Fig.~\ref{fig:2}b we see UV scaling in a broad momentum range $\tilde{t}^{\beta}k \gtrsim \SI{1}{\per\um}$, with $\alpha = -0.70(7)$ and $\beta = - 0.14(2)$.  In the inset we show the scaled $\mathcal{E}_k$ curves, which highlight variations in the UV and can be directly compared with the unscaled curves in Fig.~\ref{fig:1}c. The ratio of the scaling exponents, $\alpha/\beta \approx 5$, is consistent with energy-conserving transport; for particles with a quadratic dispersion relation in $d$ dimensions, one expects $\alpha/\beta = d$ for a particle-conserving transport and $\alpha/\beta = d + 2$ for an energy-conserving one. 

In Fig.~\ref{fig:2}c we focus on the complementary IR $k$-range, and show both unscaled (top) and scaled (bottom) distributions. Here we observe collapse for \mbox{$\tilde{t}^{\beta}k \lesssim \SI{0.5}{\per\um}$}, with $\alpha = 1.15(8)$ and $\beta = 0.34(5)$, and $\alpha/\beta \approx 3$ consistent with particle-conserving transport.


\begin{figure}[t]
   \centering
   \includegraphics[width=\columnwidth]{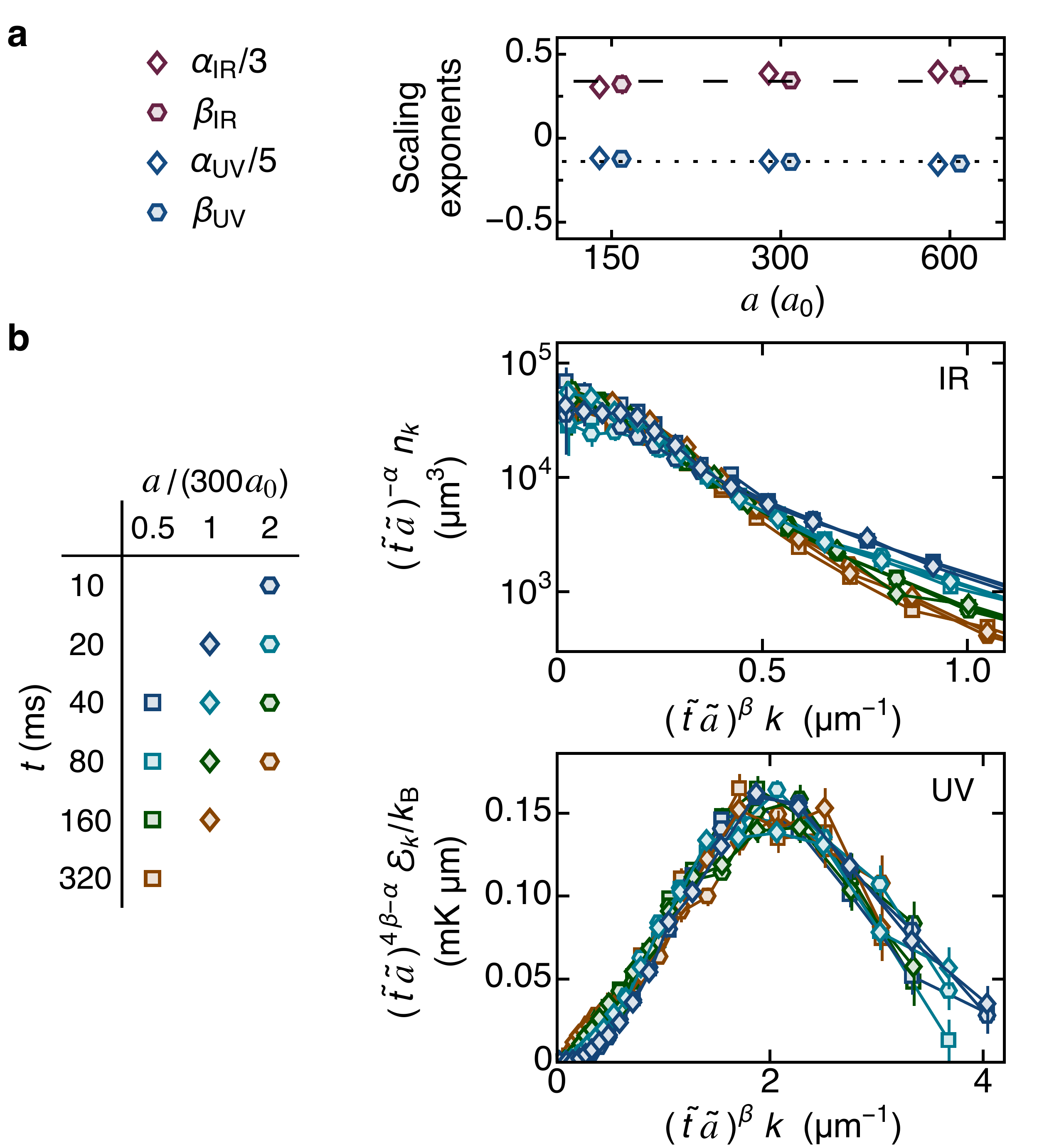}
   \caption{$\vert$ {\bf Universality for different interaction strengths.} {\bf a,}~Scaling exponents are insensitive to the value of $a$ during thermalisation; the $\alpha$ and $\beta$ points are slightly offset horizontally for visual clarity. The dashed and dotted lines, respectively, show $\beta_{\rm IR}= 0.34$ and $\beta_{\rm UV}= -0.14$.
 {\bf b,}~Generalising $t \rightarrow t \tilde{a}$ in Eq.~(\ref{eq:scaling}) collapses (separately in the IR and the UV)  all curves taken within the scaling interval $t \tilde{a} \in [20~{\rm ms},160~{\rm ms}]$; here $\tilde{a} = a/(300~a_0)$, where $300~a_0$ is an arbitrary reference point.  
 }
   \label{fig:3}
\end{figure}

We next explore the generalisation of universal dynamics to different interaction strengths (see Fig.~\ref{fig:3}), by repeating analogous experiments with $a=150~a_0$ and $600~a_0$ during thermalisation at $t>0$.
We find that all our results remain essentially the same if we rescale the thermalisation clock by $t \rightarrow t \tilde{a}$, where $\tilde{a} = a / (300~a_0)$. 
For all $a$, we observe scaling dynamics in the interaction-normalised interval $t \tilde{a} \in [20~{\rm ms},160~{\rm ms}]$, and find very similar scaling exponents, summarised in Fig.~\ref{fig:3}a; combining all our data gives $\alpha_{\rm IR} = 1.08(9)$, $\beta_{\rm IR}= 0.34(4)$,  
$\alpha_{\rm UV} = -0.67(6)$,  and $\beta_{\rm UV}= -0.14(2)$.
In Fig.~\ref{fig:3}b we show that, both in the IR and in the UV, generalising $t \rightarrow t \tilde{a}$ in Eq.~(\ref{eq:scaling}) collapses all our different-$a$ curves taken within the scaling $t \tilde{a}$-interval; here we use our $a$-averaged scaling exponents, and for visual clarity in the UV we show scaled $\mathcal{E}_k$ curves~\footnote{We have also considered a more general interaction-scaling $t \propto  a^{-p}$ and optimised the collapse of the curves in Fig.~\ref{fig:3}b with respect to $p$; this gave $p_{\rm IR}=0.9(1)$ and $p_{\rm UV}=1.1(1)$.}.
The $1/a$ scaling of the characteristic timescales implies that they are set by the inverse interaction energy, rather than the inverse two-body scattering rate, $\propto 1/a^2$; see Ref.~\cite{Davis:2017} for an overview of long-standing discussions on this issue.


\begin{figure} [tbp]
   \centering
   \includegraphics[width=\columnwidth]{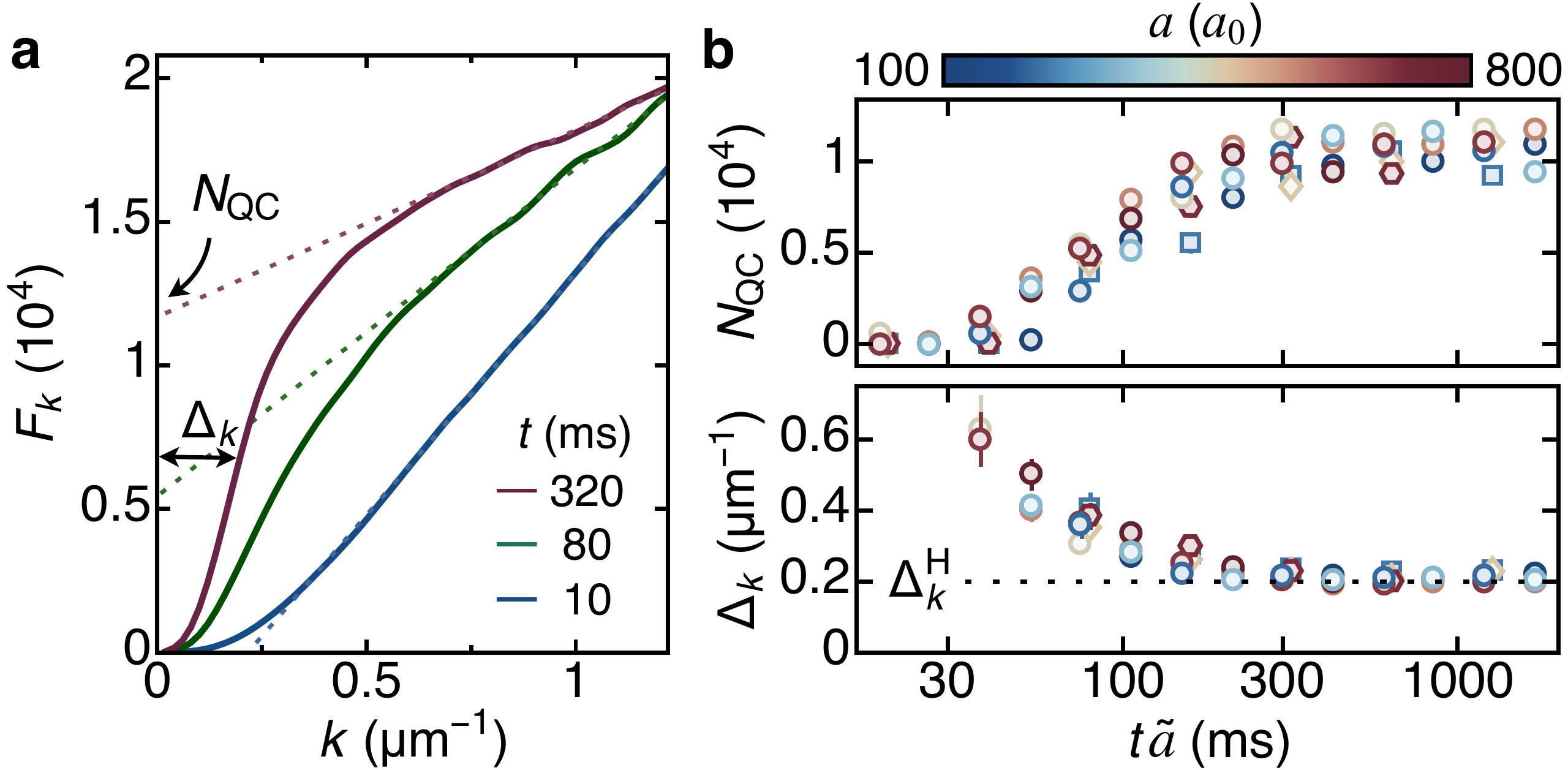}
   \caption{$\vert$ {\bf Quasi-condensation and phase ordering.} {\bf a,}~We extract the quasi-condensate atom number, $N_{\rm QC}$, and momentum-space width, $\Delta_k$, from the cumulative particle distribution $F_k$ (see text); here $a=300~a_0$. {\bf b,}~Evolution of $N_{\rm QC}$ and $\Delta_k$ for various interaction strengths is given by universal curves when plotting versus $t \tilde{a}$. At long times $\Delta_k$ approaches the Heisenberg limit (dotted line), corresponding to a fully coherent condensate.}
   \label{fig:4}
\end{figure}

Finally, we look at the properties of the condensate that emerges during thermalisation. The IR scaling in Figs.~\ref{fig:2} and~\ref{fig:3}, seen for all low $k$, implies that the momentum width of the condensate, $\Delta_k$, is shrinking. This is not consistent with a fully coherent BEC, which has a fixed Heisenberg-limited width $\Delta_k^{\rm H}$ set by the system size~\cite{Gotlibovych:2014}, with $\Delta_k^{\rm H} \rightarrow 0$ in the thermodynamic limit.  
However, it is consistent with the emergence of an out-of-equilibrium quasi-condensate (QC) that is initially riddled with excitations such as vortex loops~\cite{Berloff:2002,Weiler:2008} or domain walls~\cite{Navon:2015}, and has $\Delta_k > \Delta_k^{\rm H}$. 
Generally, $\Delta_k^{-1}$ is a characteristic coherence length (limited by system size), and only with full phase-ordering $ \Delta_k \rightarrow \Delta_k^{\rm H}$.

Inspired by Ref.~\cite{Berloff:2002}, we define the quasi-condensate atom number, $N_{\rm QC}$, and width, $\Delta_k$, as illustrated in Fig.~\ref{fig:4}a. Here, $F_k(k)=\int_0^{k}\mathcal{N}_k (k')\text{d}k'$ is the cumulative atom distribution. In the thermodynamic limit, for an ideal equilibrium gas of (large) volume $V$ and with $N_0$ condensed atoms, $F_k =N_0 + c \, Tk$ for $k\rightarrow 0$, where $c = Vm k_{\rm B} /(\pi^2 \hbar^2)$; in this case the BEC is localised in $k=0$ and the low-$k$ contribution of the saturated thermal gas to $F_k$ is $\propto T k$ because in the classical-field regime $\mathcal{N}_k \propto T$ is $k$-independent. In a finite-size and/or nonequilibrium gas, the (quasi-)condensate contribution to $F_k$ is spread over $\sim \Delta_k$, but in its presence one can still see a low-$k$ `shoulder' in $F_k$ and the linear regime at slightly larger $k$ (around $\SI{1}{\per\um}$ in Fig.~\ref{fig:4}a). We linearly fit the data for  $k > \SI{0.8}{\per\um}$ (dotted lines) and define $N_{\rm QC}$ by the positive intercept of this fit, while negative intercepts mean $N_{\rm QC}=0$~\cite{Berloff:2002};  here we still assume free particles and ideal-gas thermodynamics, but note that considering the phononic nature of low-$k$ excitations gives (within experimental scatter) the same results as shown in Fig.~\ref{fig:4}b. Finally, we define $\Delta_k$ as the $k$-range containing half of $N_{\rm QC}$, which for our system gives $\Delta_k^{\rm H} \approx \SI{0.2}{\per\um}$.

In Fig.~\ref{fig:4}b we show how $N_{\rm QC}$ and $\Delta_k$ evolve and eventually, at times beyond the scaling interval $t \tilde{a} \in [20~{\rm ms},160~{\rm ms}]$, approach their equilibrium values. Here we include additional data taken for various $a$ in the range $(100 - 800)~a_0$, which all fall onto universal curves when plotted versus $t \tilde{a}$. The QC emerges soon after the start of thermalisation, since our pre-quench gas is close to condensation, but initially $\Delta_k$ is notably above the Heisenberg limit. 
At long times, the condensed fraction $N_{\rm QC}/N$ approaches $40(5)\%$, consistent with the conserved $N$ and $E$ shown in Fig.~\ref{fig:1}b, while $\Delta_k$ approaches the Heisenberg limit, corresponding to a fully coherent BEC.


Our experiments provide a comprehensive picture of the universal bidirectional dynamic scaling in an isolated quantum gas, quasi-condensation, and phase ordering. They also raise questions for further theoretical and experimental work. 
The observed ratios of scaling exponents, $\alpha/\beta$, confirm the expectations linked to fundamental conservation laws. On the other hand, the values of the individual exponents are still subject of extensive theoretical work, for which our experiments provide invaluable benchmarks. For the UV dynamics, our $\beta_{\rm UV}=-0.14(2)$  is close to the prediction for weak-wave turbulence, 
$\beta_{\rm UV} = -1/6$~\cite{Zakharov:1992, Dyachenko:1992}.
For the IR dynamics, NTFP theories generally predict $\beta_{\rm IR}=1/2$~\cite{PineiroOrioli:2015,Berges:2015b,Chantesana:2019, Mikheev:2019, Schmied:2019b}, but recent work also suggests the possibility of  $\beta_{\rm IR}=1/3$~\cite{Mikheev:2019,Schmied:2019b}, closer to our $\beta_{\rm IR}=0.34(6)$; in the future it would be interesting to explore the conditions under which either might be observed.
Finally, it would also be interesting to perform similar quench experiments starting far above $T_{\rm c}$, since the dynamics on the way to (quasi-)condensation and following its onset are expected to be different~\cite{Davis:2017}.

We thank J. Berges, T. Gasenzer, J. Schmiedmayer, M.~K. Oberthaler, E.~A. Cornell, V. Kasper, and N. Navon for discussions. This work was supported by EPSRC [Grants No.~EP/N011759/1 and No.~EP/P009565/1], ERC (QBox), and a QuantERA grant (NAQUAS, EPSRC Grant No.~EP/R043396/1). C.~E. acknowledges support from Jesus College (Cambridge). T.~A.~H. acknowledges support from the EU Marie Sk\l{}odowska-Curie program [Grant No.~MSCA-IF- 2018 840081]. R.~P.~S acknowledges support from the Royal Society. Z.~H. acknowledges support from the Royal Society Wolfson Fellowship.

%

\vspace{9mm}

\setcounter{figure}{0} 
\setcounter{equation}{0} 

\renewcommand\theequation{S\arabic{equation}} 
\renewcommand\thefigure{S\arabic{figure}}

\section{Methods}

\textbf{Momentum distributions.} 
We take absorption images of our clouds, after a time-of-flight (ToF) ballistic expansion of variable duration $t_{\rm ToF}$, along the symmetry axis of our cylindrical box trap. For $2\hbar k t_{\rm ToF}/m$ sufficiently larger than $D$ and $L$, such images faithfully give the line-of-sight-integrated momentum distribution. To deduce $n_k$ values that vary over 5 orders of magnitude (see Fig.~\ref{fig:2})  we combine data taken with various $t_{\rm ToF}$ in the range [\SI{10}{\milli\second}, \SI{80}{\milli\second}]; the longest $t_{\rm ToF}$ is needed to minimise finite-size effects at low $k$, while shorter ones gives better signal-to-noise at large $k$. We always repeat measurements about 10 times under identical experimental conditions. 

To reconstruct the 3D momentum distribution, we average our images azimuthally and perform the inverse-Abel transform. This assumes spherical symmetry, and in the paper we always treat $n_k$ as dependent only on $k = |{\bf k}|$. For a fully coherent BEC the momentum distribution is not isotropic (but depends on the box shape), but this does not invalidate our definition of $\Delta_k$ in terms of the cumulative distribution $F_k$.

\vspace{3mm}

\textbf{Scaling exponents.} We determine the optimal $\alpha$ and $\beta$ using an $F$--statistic approach.
For a given $\{\alpha, \beta\}$ pair, we calculate the variation between the scaled $n_k$ curves taken for different $t$ within the scaling interval, focusing on the relevant (IR or UV) momentum range.
We compare this spread to the average experimental spread in the data taken for individual $t$ values and find  $\{\alpha, \beta\}$ that minimise the ratio of the two. We estimate the uncertainty in $\alpha$ and $\beta$ by sampling 80 data subsets, each containing one third of the data, and use the spread of the obtained exponents.

\end{document}